\documentclass[11pt, oneside]{article}   	
\usepackage[T1]{fontenc}
\usepackage{lmodern}
\usepackage{geometry}                		
\usepackage{graphicx}				
\usepackage{amsmath}		
\usepackage{amssymb,amsfonts,amsthm}
\usepackage{mathtools,cool}
\usepackage{bbold}
\usepackage{braket}

\title{\textbf{Extended Massive Ambitwistor String}}
\author{Christian Kunz \\ \small{\textit{E-mail:} \href{mailto:kunz.christian.321@gmail.com}{kunz.christian.321@gmail.com}}}
\usepackage{plain}
\usepackage[dvipsnames]{xcolor}
\usepackage{hyperref}
\hypersetup{
    colorlinks = true,
    linkcolor = {Fuchsia},
    citecolor = {blue},
    urlcolor={blue},
}
\usepackage{cleveref}

\newcommand{\ud}{\mathrm{d}}
\numberwithin{equation}{section}
\allowdisplaybreaks
\begin{document}
  \maketitle
  \tableofcontents

\begin{abstract}
  This work considers a variation of the massive ambitwistor string model presented in {\href{https://arxiv.org/abs/2301.11227}{arXiv:2301.11227}} that describes supergravity and super-Yang-Mills on the Coulomb branch simultaneously with a single Lagrangian. All-multiplicity tree and one-loop amplitudes are evaluated and shown to have proper unitary factorization. The massless limit provides Einstein Yang-Mills amplitudes including multiple gluon traces. It is argued that the cosmological constant vanishes at all orders of perturbation theory. As application, results for Compton scattering are obtained and compared with ones in the literature.\\
\end{abstract}
\clearpage

\section{Introduction}
There has been a lot of progress in calculating on-shell scattering amplitudes for massless and massive particles using the spinor-helicity formalism and generalized unitarity \cite{Craig:2011ws,Arkani-Hamed:2017jhn,Cachazo:2018hqa,Bern:2011qt,Wen:2020qrj}. There also has been some progress in deriving such amplitudes directly from a massive ambitwistor string theory centered on polarized scattering equations \cite{Geyer:2018, Albonico:2020, Geyer:2020, Albonico:2022, Albonico:2023}. What has been still missing, was a consolidation of supergravity and super-Yang-Mills (SYM) on the Coulomb branch into a unified theory that exhibits proper unitary factorization justifying the use of generalized unitarity. This article is a step forward in this direction.\\

The paper is organized as follows:\\

In section \ref{ExMassATStr} the supergravity version of the massive ambitwistor string presented in \cite{Albonico:2022, Albonico:2023} is extended to include additional gauge symmetries. It has some similarity with the supergravity model of Skinner \cite{Skinner:2013} and is shown to be anomaly-free in the same manner as the original massive ambitwistor string. It also has the same $\mathcal{N} \!=\! 8$ supergravity spectrum. The all-multiplicity tree amplitude is calculated.\\

In subsection \ref{consChk} some consistency checks are performed. First a massive Einstein-Yang-Mills (EYM) amplitude is derived which includes SYM on the Coulomb branch as special case and has the proper massless limit \cite{Adamo:2015}. Also the massless limit of the general tree amplitude is shown to encompass known supergravity amplitudes \cite{Geyer:2014, Albonico:2020}.\\

In subsection \ref{TreeFactorization} proper factorization of tree amplitudes is established.\\

In section \ref{Loops} one-loop amplitudes are examined and evaluated.\\

In subsection \ref{NonSeparating} unitary factorization of the amplitude is demonstrated for non-sepa\-rating degeneration of genus 1 Riemann surfaces.\\

In subsection \ref{Separating} the same is done for separating degeneration.\\

In subsection \ref{hLoops} some remarks concerning higher loop amplitudes are made. It is argued that the cosmological constant is zero at all orders of perturbation theory. Further, it is mentioned that the original ambitwistor model shares many of the features of the extended model, in particular with regards to even spin structures and the cosmological constant.\\ 

Section \ref{massiveCompton} serves as another consistency check by applying the model to the Compton scattering of a massless particle hitting a massive target. A difference to the original massive ambitwistor string for supergravity is that the same-helicity amplitude does not vanish for 'gravitinos', and the result for 'gluons' is consistent with the one arising from SYM on the Coulomb branch.\\

Section \ref{Summary} contains summary and discussion.\\

\section{Extended Massive Ambitwistor String}
  \label{ExMassATStr}
  In this section the massive ambitwistor string of \cite{Albonico:2023} for maximal supergravity is extended with an enlarged supersymmetric gauge symmetry. It can also be viewed as generalization of a massless ambitwistor string model in \cite{Kunz:2023xjj} to include massive particles.\\

The twistor action used here is, in similar notation to \cite{Kunz:2023xjj},
 \begin{align}
  \label{massSugraAction}
  S \;\,\!=\! \int_\Sigma &\mathcal{Z}^a \cdot \bar\partial \mathcal{Z}_a + A_{ab} \mathcal{Z}^a \cdot \mathcal{Z}^b + a \!\braket{\lambda_A \lambda_B}\!\Omega^{AB} + \tilde{a} (\!\braket{\lambda_A \lambda_B}\!\Omega^{AB} \!+\!\braket{\eta_\mathcal{I} \eta_\mathcal{J}}\!\Omega^{\mathcal{I}\mathcal{J}})\nonumber\\
  & + S_{\rho_1} + S_{(\rho_2, \tau)} ,\nonumber\\   
  S_{\rho}\,\!=\!\int_{\Sigma} & \frac{1}{2} \braket{\rho_A \bar{\partial} \rho_B}\Omega^{AB} + B_{ab} \lambda^a_A \rho^b_B \Omega^{AB}  + b \braket{\lambda_A \rho_B}\Omega^{AB}\,,\\ 
  S_{(\rho, \tau)}\!=\!\int_{\Sigma} & \frac{1}{2} (\braket{\rho_A \bar{\partial} \rho_B}\!\Omega^{AB} \!+\! \braket{\tau_{\mathcal{I}} \bar{\partial} \tau_{\mathcal{J}}} \!\Omega^{\mathcal{I}\mathcal{J}}) + D_{ab} (\lambda^a_A \rho^b_B \Omega^{AB} \!+\! \eta^a_{\mathcal{I}} \tau^b_{\mathcal{J}}\!\Omega^{\mathcal{I}\mathcal{J}}) \nonumber \\
  & + d \,(\braket{\lambda_A \rho_B}\Omega^{AB} + \braket{\eta_{\mathcal{I}} \tau_{\mathcal{J}}} \Omega^{\mathcal{I}\mathcal{J}}) \,,\nonumber 
  \end{align} 
  where $a = 1,2$ is the little group index with contraction notation $\xi_a \zeta_b \varepsilon^{ab} = \xi_a \zeta^a \equiv \braket{\xi \zeta}$ and $\mathcal{Z}_a$ is a pair of supertwistor fields that are worldsheet spinors, repackaged into \textit{Dirac} supertwistors of the form
  \begin{align*}
  \mathcal{Z}\!=\!(\lambda_A,\mu^A,\eta_\iota)\!:\;&\lambda_A \!=\! (\lambda_\alpha,\tilde \lambda_{\dot\alpha}) , \mu^A \!=\! (\mu^{\dot\alpha},\tilde\mu^{\alpha}),
  \;\eta_{\iota}\!=\!(\eta_\mathcal{I}, \tilde{\eta}^\mathcal{I}), \eta_{\mathcal{I}}\!=\!(\eta_I,\tilde \eta_{\dot{I}}), \tilde{\eta}^\mathcal{I}\!=\!(\tilde \eta^{\prime I}, \eta^{\prime \dot{I}}),\nonumber \\
  &I,\dot{I} = 1 \ldots \frac{\mathcal{N}}{2} , \,\mathcal{I} = 1, \ldots, \mathcal{N}, \,\iota = 1, \ldots, 2\mathcal{N}\,, \,\mathcal{N} = 8\, ,
  \end{align*}
  where $\lambda_A$ and $\mu^A$ are Dirac spinors made up of the homogeneous chiral and antichiral components of the twistor $Z = (\lambda_\alpha, \mu^{\dot\alpha})$ and dual twistor $\tilde{Z} = (\tilde \lambda_{\dot\alpha}, \tilde\mu^{\alpha})$, and $\eta_{\iota}\!=\!(\eta_\mathcal{I}, \tilde{\eta}^\mathcal{I})$ are fermionic components with $\mathcal{I}=1,\ldots,\mathcal{N}$ as the R-symmetry index, with $\mathcal{N}=8$ for maximal supergravity. Indices are raised and lowered with skew 4 $\!\times\!$ 4 forms $\Omega^{E F}, \Omega_{EF}, \Omega^{EG} \Omega_{GF} = \delta^E_F$. The symmetric $\mathcal{N} \times \mathcal{N}$ form $\Omega^{\mathcal{I} \mathcal{J}}$ decomposes into two $\frac{\mathcal{N}}{2} \times \frac{\mathcal{N}}{2}$ forms: $\Omega^{\mathcal{I} \mathcal{J}} = \begin{pmatrix}
  0 &\mathbb{1}^{I \dot{J}}\\
  \mathbb{1}_{\dot{I} J} &0 \end{pmatrix}$.
  The inner product is defined as\\
  $\mathcal{Z}_a \cdot \mathcal{Z}_b = \frac{1}{2}(\tilde{Z_a} \cdot Z_b + \tilde{Z_b} \cdot Z_a + \tilde\eta_{a\mathcal{I}} \eta^{\mathcal{I}}_b + \tilde\eta_{b\mathcal{I}} \eta^{\mathcal{I}}_a),\;\; \tilde{Z_a} \cdot Z_b = \tilde\mu_a^{\alpha}\lambda_{b \alpha} + \tilde \lambda_{a \dot\alpha} \mu_b^{\dot\alpha}$\,,\\
  with special treatment of $\bar\partial Z$ when taking the dual: $\;\widetilde{\bar\partial Z} = - \bar\partial \tilde{Z}\,$.\\
  $\rho^a_{rA} (r \!=\! 1,2)$ and $\tau^a_\mathcal{I}$ are auxiliary fermionic and bosonic worldsheet spinors, respectively, and $(B_{ab}, b, D_{ab}, d)$ are fermionic Lagrange multipliers for the constraints \\
  $\lambda^{(a}_A \rho^{b)}_{1B} \Omega^{AB} = 0 = \braket{\lambda_A \rho_{1B}}\! \Omega^{AB}, \lambda^{(a}_A \rho^{b)}_{2B} \Omega^{AB} \!+\! \eta^{(a}_{\mathcal{I}} \tau^{b)}_{\mathcal{J}}\Omega^{\mathcal{I}\mathcal{J}} = 0 = \braket{\lambda_A \rho_{2B}}\!\Omega^{AB} + \braket{\eta_{\mathcal{I}} \tau_{\mathcal{J}}} \!\Omega^{\mathcal{I}\mathcal{J}}\!$\\
   on the supersymmetric gauge currents.\\
  Little group transformations for the twistors are gauged by the fields $A_{ab} = A_{(ab)}$, and $a$ and $\tilde{a}$ are worldsheet $(0,1)$-forms that also act as Lagrange multipliers, required in order to have a closed current algebra.\\
  
  It can be noted that the action is somewhat similar to the $\mathcal{N}\!=\!8$ supergravity model of Skinner \cite{Skinner:2013} which has the same kind of auxiliary pair of supertwistors with reverse statistics as in $S_{(\rho, \tau)}$, but only a single main supertwistor. The action is even more similar to the massive ambitwistor model of \cite{Albonico:2022, Albonico:2023} and can be viewed as an extension of it. The difference consists in doubling the fermionic components in the main supertwistor, gauging them with auxiliary $\tau$ fields using the same worldsheet supersymmetry as for $\rho_2$, and replacing the constraints on mass operators with constraints on currents that close the current algebra.\\
  
  BRST quantization can be performed similarly to \cite{Geyer:2020}. In addition to the familiar fermionic ($b,c$) ghost system related to worldsheet gravity (the action \eqref{massSugraAction} is already written in conformal gauge), the Lagrange multipliers in \eqref{massSugraAction} are associated with corresponding (anti-ghost, ghost) pairs:\\
  the bosonic fields $\{A_{ab}, a, \tilde{a}\}$ with fermionic ghosts $\{(M^{ab}, N_{ab}), (m, n), (\tilde{m}, \tilde{n})\}$ and\\
  the fermionic fields $\{B_{ab}, b, D_{ab}, d\}$ with bosonic ghosts $\{(\beta^{ab}, \!\gamma_{ab}),(\beta, \!\gamma), (\tilde{\beta}^{ab}, \!\tilde{\gamma}_{ab}), (\tilde{\beta}, \tilde{\gamma})\}$.\\
  All the Lagrange multipliers can be gauged to zero and the BRST operator becomes $Q= \oint \ud z J_{BRST}(z)$, where the BRST current $J_{BRST}(z)$ is (based on the formula $Q = c^i K_i - \frac{1}{2} f^{ij}_k c^i c^j b_k$ for the current algebra $K_i$ with structure constants $f^{ij}_k$, ghosts $c^i$, and anti-ghosts $b_k$) :
 \begin{align*}
  J_{BRST} \!= &\, c\,T \!+\! N_{ab}(J^{ab} \!+\! M^a_c N^{bc}) \!+\! n \!\braket{\lambda_A \lambda^A} \!+\! \tilde{n} (\!\braket{\lambda_A \lambda^A}\!+\!\braket{\eta_{\mathcal{I}} \eta^{\mathcal{I}}})
   \!+\! \gamma_{ab} \lambda^{aA}\rho^b_{1A} \!+\! \gamma\!\braket{\lambda^A \rho_{1A}}  \nonumber\\ 
  &+\! m \gamma \gamma \!+\! \tilde{\gamma}_{ab} (\lambda^{aA}\rho^b_{2A} \!+\! \eta^{a\mathcal{I}} \tau^b_{\mathcal{I}}) \!+\! \tilde{\gamma} (\braket{\lambda^A \rho_{2A}}\!+\!\braket{\eta^{\mathcal{I}} \tau_{\mathcal{I}}}) \!+\! \tilde{m} \tilde{\gamma}\tilde{\gamma}\,,\\ 
  T \!= &\mathcal{Z}^a \!\!\cdot\! \partial \mathcal{Z}_a \!+\! \frac{1}{2}\!\sum_{r = 1,2} \!\!\braket{\rho_{rA} \partial \rho_r^A} \!+\! \beta^{ab} \partial \gamma_{ab} \!+\!\beta \partial \gamma \!+\! \frac{1}{2} \!\braket{\tau_{\mathcal{I}} \partial \tau^{\mathcal{I}}} \!+\! \tilde{\beta}^{ab} \partial \tilde{\gamma}_{ab} \!+\! \tilde{\beta} \partial \tilde{\gamma} \nonumber\\
   & +\! M^{ab} \partial N_{ab} + m \partial n \!+\! \tilde{m} \partial \tilde{n} \!+\! \frac{1}{2}(b \partial c + \partial(bc))\,, \nonumber\\
  J^{ab} \!= & \mathcal{Z}^a \cdot \mathcal{Z}^b  \!+\! \frac{1}{2}\!\sum_{r = 1,2} \rho^{(a}_{rA}  \rho_r^{b)A} \!+\! \beta^{c(a}  \gamma^{b)}_c \!+\! \frac{1}{2} \tau^{(a}_{\mathcal{I}} \tau^{b)\mathcal{I}}  \!+\! \tilde{\beta}^{c(a}  \tilde{\gamma}_c^{b)} \,,\nonumber
  \end{align*}
  where $T$ is the energy-momentum current and $J^{ab}$ the sl(2, $\mathbb{C}$) current. Obstructions to the nilpotency of the BRST charge $Q$ can only come from the sl(2,$\mathbb{C}$) algebra and the Virasoro algebra. The sl(2, $\mathbb{C}$) anomaly coefficient is, using $\text{tr}_{\text{adj}}(t^kt^k) = 6, \text{trf}_{\text{F}}(t^kt^k) = 3/2$,
  \begin{equation*}
    a_{sl2} = \frac{1}{2}\text{trf}_{\text{F}}(t^kt^k) ((8 \!-\! 2\mathcal{N})_{\!\mathcal{Z}} \!-\! 4_{\rho_1} \!-\! 4_{\rho_2} \!+\! \mathcal{N}_{\tau}) + \text{tr}_{\text{adj}}(t^kt^k)(\!-1_{M\! N} + 1_{\beta \gamma} + 1_{\tilde{\beta} \tilde{\gamma}})  \!=\! \frac{3}{4}(8 - \mathcal{N}) = 0\,. 
    \end{equation*}
    For the central charge it is assumed that there is a contribution of 12 from compactifying 6 dimensions like for the massive ambitwistor string model of \cite{Albonico:2023}, e.g. with help of 6 bosonic scalars as in appendix A of \cite{Albonico:2023}. Then it can be seen to vanish for $\mathcal{N}=8$:
    \begin{equation*}
    c \!=\! (\!-8 + 2 \mathcal{N}\!)_{\!\mathcal{Z}} \!- 26_{bc} \!- 6_{M\! N} \!-2_{mn} \!- 2_{\tilde{m}\tilde{n}} \!+ 4_{\rho_1} \!+ 8_{\beta \gamma} \!+ 4_{\rho_2} \!-\! \mathcal{N}_{\!\tau} + 8_{\tilde{\beta} \tilde{\gamma}}\!  + 12_{\text{comp}} \!=\! -8 + \! \mathcal{N} \!=\! 0\,.
    \end{equation*}
    Therefore, the model can be viewed as conditionally anomaly-free. The additional contribution to the central charge can also be obtained by gauging additional worldsheet supersymmetries between additional auxiliary spinors that do not affect the spectrum and vertex operators \cite{Kunz:2025}. In this manner the model can become anomaly-free without resorting to extra spacetime dimensions. Of course, there might be many more ways to come up with an auxiliary action that achieves the same outcome.\\
    
    To find the spectrum one needs to determine the Virasoro $L_0$ constant $a$ \cite{Kunz:2023xjj}. In the NS sector $a$ is calculated by changing in the calculation of the central charge the $bc$ contribution from $-26_{bc}$ to $-2_{bc}$ and dividing the result by 24, such that $a = 1$ in the NS sector. In the R sector $a$ is:
    \begin{equation*}
    24 a \!=\! \!(\!16 - \!4\mathcal{N}\!)_{\!\mathcal{Z}} \!-\! 2_{bc} \!- 6_{M\! N} \!-2_{mn} \!- 2_{\tilde{m}\tilde{n}}\!- 8_{\rho_1} \!+ 8_{\beta \gamma} \!-( 8_{\rho_2} \!-\! 2\mathcal{N}_{\!\tau}) + 8_{\tilde{\beta} \tilde{\gamma}}\!  + 12_{\text{comp}} \!=\! 2(8 - \! \mathcal{N}) \!=\! 0\,.
    \end{equation*}
    i.e. $a = 0$ in the R sector.\\
  The situation is parallel to the massive ambitwistor string model of \cite{Albonico:2023}, and based on BRST cohomology the spectrum is again the one of $\mathcal{N}\!=\!8$ supergravity with SU(4) $\!\times\!$ SU(4) $R$-symmetry. Although the fermionic modes in the supertwistors have been doubled, similarly to the exclusion of $\mu^{aA}_n$ modes from the spectrum because of BRST cohomology, now the $\tilde{\eta}^{a\mathcal{I}}_n$ cannot contribute either to the spectrum because of the requirement that the non-negative modes of the $\eta^{\mathcal{I}(a} \tau^{b)}_{\mathcal{I}}$ and $\braket{\eta^{\mathcal{I}} \tau_{\mathcal{I}}}$ currents annihilate physical states. One other difference is that the action in \cite{Albonico:2023} contains mass operators that ensure that the supersymmetric multiplet has a particular mass whereas the current model leaves the value of the mass open.\\
  
  Moving on to vertex operators, following \cite{Geyer:2020, Albonico:2022, Albonico:2023} vertex operators are sought in the form  
  \begin{equation}
  \label{VO}
  \genfrac{(}{)}{0pt}{}{\mathcal{V}}{V} \!=\!\! \int \!\ud^2\!u \ud^2\!v \genfrac{(}{)}{0pt}{}{\mathcal{W}(u)}{w(u)}  \bar{\delta}(\!\braket{v \epsilon} \!-\!1) \bar{\delta}^{4}(\!\braket{u \lambda_A} \!-\! \braket{v \kappa_A}) \bar{\delta}^{\mathcal{N}}\!(\!\braket{u \eta_{\mathcal{I}}} \!-\! \braket{v \zeta_{\mathcal{I}}}) \; e^{\braket{u \mu^A} \epsilon_A + \braket{u \tilde{\eta}^{\mathcal{I}}} q_{\mathcal{I}}}.\\
  \end{equation}
  Here $\epsilon_a$ are little group spinors, $\kappa^A_a = (\kappa^\alpha_a, \tilde{\kappa}^{\dot{\alpha}}_a)$ are momentum spinors with associated polarization spinors $\epsilon^A \!=\! \braket{\epsilon \kappa^A}$,$\;\zeta^{\mathcal{I}}_a$ are supermomenta with superpolarization spinors $q^{\mathcal{I}} \!=\! \braket{\epsilon \zeta^{\mathcal{I}}}$, and $\mathcal{V}$ and $V$ stand for fixed and integrated vertex operators, respectively, which differ only by an operator $\mathcal{W}$ or $w$ with $\mathcal{W}$ typically being just a product of fermionic ghost fields and delta functions of bosonic ghost fields. As the handling of fixed and integrated vertex operators concerning the $(b,c)$ ghosts is standard, the difference being just an integration over the worldsheet, the $(b,c)$ ghosts and worldsheet integrations will be omitted in the formulae for $\mathcal{W}$ and $w$, but then added later in the correlators. Also the effect of $(m,n)$ is just to fix $\braket{\lambda_{\alpha} \lambda^{\alpha}} = \braket{\lambda_{\dot\alpha} \lambda^{\dot\alpha}}$ which can be viewed as a little group transformation outside of sl(2, $\mathbb{C}$) \cite{Arkani-Hamed:2017jhn,Penrose:1986}  and will be skipped as well, and similarly for $(\tilde{m}, \tilde{n})$. Finally it is assumed that the bosonic fields from compactifying extra dimensions or additional supersymmetric gauge symmetries do not affect the spectrum or vertex operators\footnote{They will be crucial for loop amplitudes in the next section.}. For more details in this regard see \cite{Kunz:2025}. The remainder can be simplified even further because the form \eqref{VO} is such that it is sufficient that the delta functions of $\gamma_{ab}$ and $\tilde{\gamma}_{ab}$ in $\mathcal{W}$ force only the ghost components orthogonal to $u$ to vanish \cite{Geyer:2020}. Consequently 
  \begin{align*}
  \mathcal{W}(u) \!=& \,\delta(\gamma) \,\delta(u^a u^b \gamma_{ab}) \;\delta(\tilde{\gamma})\, \delta(u^a u^b \tilde{\gamma}_{ab}), \\
    w(u) \!=& \!\left(\frac{\braket{\hat{u} \lambda_A} \epsilon^A}{\braket{u \hat{u}}} \!+\! \frac{\!\braket{\rho_{1A} \rho_{1B}}}{2}\epsilon^A \epsilon^B\!\right)\! 
    \left(\frac{\braket{\hat{u} \lambda_A} \!\epsilon^A}{\braket{u \hat{u}}} \!+\! \frac{\!\braket{\rho_{2A} \rho_{2B}}}{2}\epsilon^A \epsilon^B \!+\! \frac{\braket{\hat{u} \eta_\mathcal{I}}\! q^\mathcal{I}}{\braket{u \hat{u}}} \!+\! \frac{\!\braket{\tau_\mathcal{I} \tau_\mathcal{J}}}{2}q^\mathcal{I} \!q^\mathcal{J}\!\right) \nonumber.
  \end{align*}
  Here $\hat{u}$ is a reference little group spinor that forms a basis with $u$ such that $\braket{u \hat{u}} \ne 0$. In amplitudes it will fall out based on solutions of the scattering equations \cite{Geyer:2020}.\\
  
The all-multiplicity tree amplitude of this model can now be evaluated:
  \begin{equation*}
  \mathcal{A}_n^{\text{tree}} = \left< c\mathcal{V}_1^{(-1,-1)}  c\mathcal{V}_2^{(-1,0)} c\mathcal{V}_3^{(-1,0)}\, \prod_{j =4}^n \int_{\Sigma} \! V_j \right>\,, 
  \label{amplitude-1}
  \end{equation*}
  where $\mathcal{V}^{(-1,-1)}$ denotes a fixed vertex operator with regards to all $(\beta^{ab},\!\gamma_{ab}), (\tilde{\beta}^{ab}, \!\tilde{\gamma}_{ab}), (\beta,\!\gamma),$ $ (\tilde{\beta}, \!\tilde{\gamma})$ ghosts and $\mathcal{V}^{(-1,0)}$ only fixed with regards to $(\beta^{ab},\gamma_{ab}), (\tilde{\beta}^{ab}, \tilde{\gamma}_{ab})$.\\
  
  Integrating out the $\mu^A$ and $\tilde{\eta}^\mathcal{I}$ leads, on the support of the delta functions, to the polarized scattering equations (using the notation $\sigma_{ij} = \sigma_i - \sigma_j$):
  \begin{equation}
  \label{scattEq}
  \begin{aligned}
  &\braket{v_r \kappa_r^A} \! = \! \braket{u_r \lambda^A(\sigma_r)}, &\lambda^a_A(\sigma_r) = \sum_{l \ne r} \frac{u^a_l \epsilon_{l A}}{\sigma_{rl}}, \\
  &\braket{v_r \zeta_r^{\mathcal{I}}} = \! \braket{ u_r \eta^{\mathcal{I}}(\sigma_r)}, &\eta^a_{\mathcal{I}}(\sigma_r) = \sum_{l \ne r} \frac{u^a_l q_{l\mathcal{I}}}{\sigma_{rl}}.
  \end{aligned}
  \end{equation}
    
  Based on these scattering equations and because of the delta functions enforcing $\braket{v_r \epsilon_r} \!=\! 1$ it follows that $\braket{u_r \lambda^A(\sigma_r)}\epsilon_{rA} = \braket{v_r \kappa_r^A}\epsilon_{rA} = 0$ and $\braket{ u_r \eta^{\mathcal{I}}(\sigma_r)} q_{r \mathcal{I}} = \braket{v_r \zeta_r^{\mathcal{I}}} q_{r \mathcal{I}} = 0$ such that the dependence on $\hat{u}$ falls out of the correlator.\\
  
   Then the amplitude is evaluated to
   \begin{equation}
   \label{treeAmpl}
   \mathcal{A}_n^{\text{tree}} = \int \ud \mu_n^{\text{pol}|\mathcal{N}} \;\mathcal{I}_n\,,
   \end{equation}
   where $\ud \mu_n^{\text{pol}|\mathcal{N}}$ is the measure which localizes on the scattering equations \eqref{scattEq} and also accounts for the quotient by the volume of $\text{SL}(2, \mathbb{C})_{\sigma} \times \text{SL}(2, \mathbb{C})_u$ from the path integral over the $(b,c)$ and $(M^{ab}, N_{ab})$ ghost systems:
  \begin{equation}
  \ud \mu_n^{\text{pol}|\mathcal{N}} \!:= \!\frac{\prod_l \ud \sigma_l \ud^2 u_l \ud^2 v_l}{\text{vol}(\text{SL}(2, \mathbb{C})_{\sigma} \!\times\! \text{SL}(2, \mathbb{C})_u)}\!
  \!\prod_{r = 1}^n \!\bar{\delta}(\!\braket{v_r \epsilon_r} \!-\!1) \bar{\delta}^{4}(\!\braket{u_r \lambda^A} \!-\! \braket{v_r \kappa_r^A}) \bar{\delta}^{\mathcal{N}}(\!\braket{u_r \eta^{\mathcal{I}}} \!-\! \braket{v_r \zeta_r^{\mathcal{I}}}),
  \label{measure}
  \end{equation}
   and where the integrand $\mathcal{I}_n$ is the product of two reduced determinants,
   \begin{equation}
   \label{Matrices}
   \begin{aligned}
   \mathcal{I}_n = &\quad\text{det}^\prime H \text{det}^\prime G \quad = &\frac{\text{det} H^{[ij]}_{[kl]}}{\braket{u_i u_j}\braket{u_k u_l}}\;\frac{\text{det} G^{[pr]}_{[st]}}{\braket{u_p u_r}\braket{u_s u_t}},\\
   &H_{ij} = \frac{\epsilon_{iA} \epsilon_j^A}{\sigma_{ij}}, &u_i^a H_{ii} = - \lambda_A^a \epsilon_i^A,\qquad\qquad\;\\
   &G_{ij} = \frac{\epsilon_{iA} \epsilon_j^A + q_{i\mathcal{I}} q_j^\mathcal{I}}{\sigma_{ij}}, &u_i^a G_{ii} = - \lambda_A^a \epsilon_i^A - \eta^a_\mathcal{I} q_i^\mathcal{I},\quad 
   \end{aligned}
   \end{equation}
   where $H$ and $G$ are $n\!\times\!n$ matrices with co-rank 2 and $H^{[ij]}_{[kl]}$ means that rows $i$ and $j$ and columns $k$ and $l$ have been removed. The reduced determinants are independent of the choice of rows and columns to remove.\\
   
   The amplitude \eqref{treeAmpl} has been worked out in the so-called little group (LG) representation \cite{Albonico:2020}, mainly because the derivation required 'scattering equations' \eqref{scattEq} for the $\eta$ fields, but in the following it is more convenient to go over to the R-symmetry preserving representation, by Fourier transforming both halves of the $q_\mathcal{I} = (q_I, \tilde{q}_{\dot{I}}) = (q_I, \!\braket{\epsilon \zeta_I}) = (\!\braket{\epsilon \tilde{\zeta}_{\dot{I}}}, \tilde{q}_{\dot{I}})$ variables, using $N \!=\! \tilde{N} \!=\! \frac{\mathcal{N}}{2}$ and $v_a \!=\! \xi_a \!+\! \braket{\xi v} \epsilon_a$ where $(\epsilon, \xi)$ form a spinor basis with $\braket{\xi \epsilon} = 1$\cite{Albonico:2020}:
   \begin{align}
   \label{Fourier}
   2^{-N} \!\int \prod_{i=1}^n \ud^N \!q_i^I &\prod_{j=1}^n e^{-\frac{1}{2}q_i^I \!\braket{\xi_i \zeta_{iI}}} e^{F_N} = \prod_{i=1}^n \delta^N \!\left(\sum_{j=1}^n \frac{\braket{u_i u_j}}{\sigma_{ij}} \!\braket{\epsilon_j \zeta_{j I}} - \braket{v_i \zeta_{j I}}\right), \nonumber\\
   &F_N = \sum_{i < j} \frac{\braket{u_i u_j}}{\sigma_{ij}}q_{iI} q_j^I - \frac{1}{2}\sum_{i = 1}^n \braket{\xi_i v_i} q_i^2,\\
   2^{-\tilde{N}} \!\int \prod_{i=1}^n \ud^{\tilde{N}} \!\tilde{q}_i^{\dot{I}} &\prod_{j=1}^n e^{-\frac{1}{2}\tilde{q}_i^{\dot{I}} \!\braket{\xi_i \tilde{\zeta}_{i\dot{I}}}} e^{F_{\tilde{N}}} = \prod_{i=1}^n \delta^{\tilde{N}} \!\left(\sum_{j=1}^n \frac{\braket{u_i u_j}}{\sigma_{ij}} \!\braket{\epsilon_j \tilde{\zeta}_{j \dot{I}}} - \braket{v_i \tilde{\zeta}_{j \dot{I}}}\right), \nonumber\\
   &\tilde{F}_{\tilde{N}} = \sum_{i < j} \frac{\braket{u_i u_j}}{\sigma_{ij}} \tilde{q}_{i\dot{I}} \tilde{q}_j^{\dot{I}}  - \frac{1}{2}\sum_{i = 1}^n \braket{\xi_i v_i} \tilde{q}_i^2.\nonumber
   \end{align}
   When doing this, the $q$ variables in the matrix $G$ and in $F_{\mathcal{N}} = F_N + \tilde{F}_{\tilde{N}}$ will have transposed halves, but because the inner products are the same one can just transpose one set of these $q$ variables, e.g. the ones in $F_{\mathcal{N}}$. One gets
   \begin{align}
   \label{treeAmpl1}
   \mathcal{A}_n^{\text{tree}} &= \int \ud \mu_n^{\text{pol}} \;\mathcal{I}_n e^{F_{\mathcal{N}}}, \qquad F_{\mathcal{N}} = \sum_{i < j} \frac{\braket{u_i u_j}}{\sigma_{ij}}q_{i\mathcal{I}} q_j^{\mathcal{I}} - \frac{1}{2}\sum_{i = 1}^n \braket{\xi_i v_i} q_i^2\,,\\
   &\ud \mu_n^{\text{pol}} \!:= \!\frac{\prod_l \ud \sigma_l \ud^2 u_l \ud^2 v_l}{\text{vol}(\text{SL}(2, \mathbb{C})_{\sigma} \!\times\! \text{SL}(2, \mathbb{C})_u)}\!
  \!\prod_{r = 1}^n \!\bar{\delta}(\!\braket{v_r \epsilon_r} \!-\!1) \bar{\delta}^{4}(\!\braket{u_r \lambda^A} \!-\! \braket{v_r \kappa_r^A}) ,\nonumber
   \end{align} 
   where compared to \eqref{measure} the measure has lost the delta function in the fermionic fields, replaced by the exponential factor exp($F_{\mathcal{N}}$).\\
   
   Now the only difference in the tree amplitude to the original one \cite{Albonico:2020, Wen:2020qrj} is that a second occurrence of det$^\prime H$ is replaced by det$^\prime G$. The elements in $G$ might look surprising as the terms containing $\epsilon^A$ and the ones containing $q^\mathcal{I}$ seem to refer to incompatible quantities, but one has to keep in mind that the amplitude in \eqref{treeAmpl1} is a sum of component amplitudes describing different physical situations, and the sum of quantities in elements from the matrix $G$ are always split up among component amplitudes. In other words, \eqref{treeAmpl1} should be viewed as a power series in the $q$ variables with the coefficients being component amplitudes describing different physics, although some of them are just permutations of each other. In terms of natural units the elements in $G$ containing $\epsilon^A$ should be viewed as multiplied with the gravitational coupling constant $\kappa \!=\! \sqrt{G_N}$, i.e. the $q^\mathcal{I}$-terms in $G$ effectively lower the power in gravitational couplings, in contrast to the terms in exp($F_{\mathcal{N}})$. One important benefit of \eqref{treeAmpl1} is that component amplitudes exist that contain a term $q_i^{\mathcal{I}} q_{j \mathcal{I}}$ even when $\braket{u_i u_j} \!=\! 0$, allowing for more realistic situations, particularly between massless particles of the same helicity when the exponential factor exp($F_{\mathcal{N}}$) cannot contribute because of $\braket{u_i u_j} \!=\! 0$.\\
   
  \subsection{Consistency Checks: Einstein-Yang-Mills Amplitudes and Massless Limit}
  \label{consChk}
  The purpose of this subsection is to check whether some component amplitudes are compatible with known amplitudes that describe interesting physics. One example are Einstein-Yang-Mills (EYM) amplitudes.\\
  
  First it is to be noted that all off-diagonal elements in the $G$ matrix come from the auxiliary spinor fields and only the diagonal elements reflect contractions of the main supertwistors that are relevant for physical interpretation. Collect all diagonal elements with non-zero $q$-dependence and consider all component amplitudes that contain the product of these diagonal elements with the contributions from $\lambda_A$ removed and all other component amplitudes with the same (maximum) power in $q$ variables from det$^{\prime} G$. Consider the case where all $q$-products are connected to each other to form a single trace (loop) of length $m$. Then the component amplitude arises from the determinant of the product of two matrices $\hat{H}$ and $\hat{G}$, with $(n\!-\!m)\!\times\!(n\!-\!m)$ matrix $\hat{H}$ having all rows and columns containing a $q$ dependence removed and $m\!\times\!m$ matrix $\hat{G}$ containing only the $q$ dependent elements that have been removed from $\hat{H}$. Obviously $\hat{G}$ inherits co-rank 2 from the original matrix $G$, and $\hat{H}$ is no longer singular. Therefore, for a single trace amplitude:
  \begin{align}
   \label{1TraceAmpl}
   \mathcal{A}_{n, 1 \text{tr}}^{\text{tree}} &= \int \ud \mu_n^{\text{pol}} \frac{\text{det} H^{[ij]}_{[kl]}}{\braket{u_i u_j}\braket{u_k u_l}}\;\text{det} \hat{H} \frac{\text{det} \hat{G}^{[pr]}_{[st]}}{\braket{u_p u_r}\braket{u_s u_t}} e^{F_{\mathcal{N}}}.
   \end{align} 
 By pulling down $\frac{\braket{u_p u_r}}{\sigma_{pr}} q_{p\mathcal{I}} q_r^{\mathcal{I}}$ and $\frac{\braket{u_s u_t}}{\sigma_{st}} q_{s\mathcal{I}} q_t^{\mathcal{I}}$ from the exponential factor $F_\mathcal{N}$ to close the trace (also recognizing that the term $\braket{\xi_i v_i} q_i^2$ in $F_{\mathcal{N}}$ vanishes for a particle referred to by a singly indexed $q$ and does not contribute) one gets a Parke-Taylor(PT) factor with $q$-dependent numerator $N(q)$:
 \begin{align}
   \label{1TraceAmpl1}
   \mathcal{A}_{n, 1 \text{tr}}^{\text{tree}} &= \int \ud \mu_n^{\text{pol}} \text{det}^{\prime} H\;\text{det} \hat{H} \,\text{PT} \,e^{F_{\mathcal{N}}}, \quad\text{PT} \!=\! \sum_{\text{Perm}}\frac{N(q)}{\sigma_{m1}\prod_{i=1}^{m-1} \sigma_{i\,i+1}}.
   \end{align}
  This is the single trace 'massive EYM' amplitude with no restrictions on the individual masses. When $m=n$, $\hat{H}$ is empty and \eqref{1TraceAmpl1} is then a SYM amplitude on the Coulomb branch \cite{Albonico:2020, Albonico:2023}, i.e. the current model covers both supergravity and SYM. But it should be noted that by being embedded in $\mathcal{N}=8$ supergravity this SYM has more supersymmetry than the conventional SYM.\\
  
  Concerning multiple traces the same procedure of removing all non-zero $q$-dependent elements in the matrix $G$ still would result in a degeneration of $G$ into a product of two matrices $\hat{H}$ and $\hat{G}$, but $\hat{G}$ would have a co-rank of twice the number of traces. Therefore, instead of removing all non-zero $q$-dependent elements in $G$ one should keep 2 rows and 2 columns for all traces except one and sum over all ways of doing this. In each summand $\hat{H}$ gets all $q$-dependence removed and is then a non-singular $t\!\times\!t$ matrix with $t \!=\! n \!-\! 2 \!-\! \sum_k (l_k \!-\! 2)$, where $l_k$ is the length of trace $k$, whereas $\hat{G}$ again has co-rank 2. To get a PT factor for each trace  one needs to pull down two elements from the exponential factor $F_\mathcal{N}$ for each trace corresponding either to the two rows and columns kept in $\hat{H}$ or to the two rows and columns removed from $\hat{G}$ like for the single trace.\\
  
  Another consistency check is to take the massless limit. This is a trivial generalization of section 6 in \cite{Albonico:2020}. The amplitude for $k$ negative helicity and $n\!-\!k$ positive helicity massless particles becomes, using the standard notation $\braket{\epsilon_i \epsilon_j} \!=\! \varepsilon^{\alpha \beta} \epsilon_{i \alpha} \epsilon_{j \beta} \!=\! -\!\braket{\epsilon_j \epsilon_i}, [\epsilon_i \epsilon_j] \!=\! \varepsilon^{\dot\alpha \dot\beta} \epsilon_{i \dot\alpha} \epsilon_{j \dot\beta} \!=\! -[\epsilon_j \epsilon_i]$ (not to be confused with similar usage of $\braket{}$ for little group contractions)
  \begin{align}
   \label{treeAmpl4d}
   \mathcal{A}_{n,k}^{4d,\text{tree}} &= \int \ud \mu^{4d}_{n,k} \;\text{det}^{\prime} G_{-} \;\text{det}^{\prime} G_{+} \;e^{F^k_{\mathcal{N}}}, \qquad \text{det}^{\prime}G_{\pm} = \frac{\text{det}G_{\pm}^{[st]}}{u_s u_t} \,,\\
   &\ud \mu^{4d}_{n,k} \!:= \!\frac{\prod_{l=1}^n \ud \sigma_l \ud u_l / u_l}{\text{vol}(\text{GL}(2, \mathbb{C}))}\!
  \!\prod_{i = 1}^k \bar{\delta}^{2}(u_i \tilde{\lambda}_{\dot\alpha}(\sigma_i) \!-\! \frac{ \tilde{\kappa}_{i \dot\alpha}}{\epsilon_i}) \!\prod_{p = k+1}^n \bar{\delta}^{2}(u_p \lambda_{\alpha}(\sigma_p)\!-\! \frac{ \kappa_{p \alpha}}{\tilde{\epsilon}_p}), \nonumber\\
  &F^k_{\mathcal{N}} = \sum^k_{i =1}\sum^n_{j=k+1} \frac{u_i u_j}{\sigma_{ij}}q_{i\mathcal{I}} q_j^{\mathcal{I}},\quad \lambda_{\alpha}(\sigma) \!=\! \sum_{i=1}^k \frac{u_i \epsilon_{i\alpha}}{\sigma \!-\! \sigma_i}, \quad \tilde{\lambda}_{\dot\alpha}(\sigma) \!=\! \sum_{i=k+1}^n \frac{u_i \epsilon_{i \dot\alpha}}{\sigma \!-\! \sigma_i},\nonumber\\
  &\qquad\qquad\qquad\qquad\qquad\qquad\;\;\, \eta_I(\sigma) \!=\! \sum_{i=1}^k \frac{u_i q_{i I}}{\sigma \!-\! \sigma_i}, \quad \tilde{\eta}_{\dot{I}}(\sigma) \!=\! \sum_{i=k+1}^n \frac{u_i q_{i \dot{I}}}{\sigma \!-\! \sigma_i},\nonumber\\
  &G_{-}^{ij} = \frac{\braket{\epsilon_i \epsilon_j} \!+\! q_{i I} q_j^I}{\sigma_{ij}}, \quad G_{-}^{ii} = - \frac{\braket{\epsilon_i \lambda(\sigma_i)} + q_i^I \eta_{I}(\sigma_i)}{u_i},\quad 1 \leq i,j \leq k,\nonumber\\
  &G_{+}^{ij} = \frac{[\tilde{\epsilon}_i \tilde{\epsilon}_j] \!+\! q_{i \dot{I}} q_j^{\dot{I}}}{\sigma_{ij}}, \,\quad G_{+}^{ii} = - \frac{[\tilde{\epsilon}_i \tilde{\lambda}(\sigma_i)] \,+ q_i^{\dot{I}} \tilde{\eta}_{\dot{I}}(\sigma_i)}{u_i},\quad k+1 \leq i,j \leq n,\nonumber
   \end{align}
   where $G^{[st]}$ denotes a matrix $G$ with row $s$ and column $t$ removed. Up to the additional entries in the matrices coming from fermionic fields, this is the tree amplitude of \cite{Geyer:2014, Albonico:2020}. The direct massless limit of the single trace EYM amplitude \eqref{1TraceAmpl1} is obtained similarly: 
    \begin{equation*}
   \label{1TraceAmpl4d}
   \mathcal{A}_{n,k,1 \text{tr}}^{4d,\text{tree}} = \int \ud \mu^{4d}_{n,k} \;\text{det} \hat{H}_{-} \text{det} \hat{H}_{+} \;\text{PT}\;e^{F^k_{\mathcal{N}}},
   \end{equation*}
   where $\hat{H}_{\pm}$ is the matrix obtained from $G_{\pm}$ by removing all rows and columns containing non-vanishing fermionic entries. This is exactly the single trace EYM amplitude of \cite{Adamo:2015} if there the reduced determinants are calculated by removing the last row and column representing 'squeezed' gluons. It is interesting to notice that trying to get EYM amplitudes as component amplitudes directly from \eqref{treeAmpl4d} runs into the limitation that all positive helicity gluons must be adjacent to each other. The same obstruction occurred in the massless models of \cite{Kunz:2023xjj}.\\
   
   \subsection{Factorization of Tree Amplitudes}
   \label{TreeFactorization}
   Again this becomes a straightforward exercise by building upon work done in section 7 of \cite{Albonico:2020}. There it has been shown that the measure $\ud \mu_n^{\text{pol}}$, the exponential factor $e^{F_\mathcal{N}}$, and the reduced determinants det$^\prime H$ without fermionic elements factorize correctly. Therefore, what remains to be shown is proper factorization of the reduced determinant of $G$ with fermionic contributions included. One observation is that the matrix elements of $G$ can be rewritten as, when using the 'super' notation $\epsilon_j^{\mathcal{A}} = (\epsilon_j^A, q_i^{\mathcal{I}}), \zeta^a_{\mathcal{A}} = (\lambda^a_A, \eta^a_{\mathcal{I}})$:
   \begin{equation*}
   G_{ij} = \frac{\epsilon_{i\mathcal{A}} \epsilon_j^{\mathcal{A}}}{\sigma_{ij}},\quad u^a_i G_{ii} = - \zeta^a_{\mathcal{A}} \epsilon_i^{\mathcal{A}}, 
   \end{equation*} 
   with the 'super' spinor fulfilling the polarized scattering equations \eqref{scattEq} in unified notation
   \begin{equation*}
   \braket{v_i \kappa_i^{\mathcal{A}}} = \braket{u_i \zeta^{\mathcal{A}}(\sigma_i)}, \quad \zeta^a_{\mathcal{A}}(\sigma_i) = \sum_{j \ne i} \frac{u^a_j \epsilon_{j\mathcal{A}}}{\sigma_{ij}},
   \end{equation*} 
   where $\kappa_i^{\mathcal{A}} = (\kappa_i^A, \zeta_i^{\mathcal{I}})$. Then the proof of factorization of the reduced determinants in \cite{Albonico:2020} can be repeated basically word by word by replacing $(\epsilon^A, \lambda_A, \kappa^A)$ with $(\epsilon^{\mathcal{A}} , \zeta_{\mathcal{A}} , \kappa^{\mathcal{A}} )$.\\
   
   More effort will be spent on the proof of factorization at the one-loop level in the next section.\\

  \section{Loop Amplitudes and Factorization}
  \label{Loops}
  When calculating multi-loop amplitudes in QFT, this is typically done assuming generalized unitarity \cite{Bern:2011qt}. Therefore, in this work using a model with a Lagrangian, consistency requires that this model exhibits proper factorization of scattering amplitudes to justify using generalized unitarity.\\
  
  Before evaluating one-loop amplitudes it is interesting to remark on vacuum partition functions. They vanish on even spin structures because of spinor fields that cannot get contracted, and on odd spin structures there are fermionic zero modes that cannot be cancelled or saturated. More details can be found in \cite{Kunz:2025}.\\
  
  Moving on to the one-loop scattering amplitudes themselves, the scattering equations \eqref{scattEq} change to 
  \begin{equation}
  \label{scattEq1L}
  \begin{aligned}
  &\braket{v_r \kappa_r^A} \! = \! \braket{u_r \lambda^A(\sigma_r)}, &\lambda^a_A(\sigma_r) = \lambda^a_{0A} + \sum_{l \ne r} u^a_l \epsilon_{l A} \frac{S_{\alpha}(\sigma_r, \sigma_l; \tau)}{\sqrt{\ud \sigma_r}\sqrt{\ud \sigma_l}}, \\
  &\braket{v_r \zeta_r^{\mathcal{I}}} = \! \braket{ u_r \eta^{\mathcal{I}}(\sigma_r)}, &\eta^a_{\mathcal{I}}(\sigma_r) = \eta^a_{0 \mathcal{I}} +  \sum_{l \ne r} u^a_l q_{l\mathcal{I}} \frac{S_{\alpha}(\sigma_r, \sigma_l; \tau)}{\sqrt{\ud \sigma_r}\sqrt{\ud \sigma_l}},\;\;
  \end{aligned}
  \end{equation}
  where $\lambda^a_{0A}$ and $\eta^a_{0 \mathcal{I}}$ are zero modes on an odd spin structure $\alpha = 1$ but generally do not exist on even spin structures $\alpha = 2,3,4$, and $S_\alpha(\sigma, \tau)$ is the Szeg\"{o} kernel for spin structure $\alpha$. Zero modes of the auxiliary fields $\rho^a_{1,2}$ and $\tau^a$ do not appear as these fields do not contribute to the spectrum and, therefore, are only permitted as virtual fields, so their zero modes can be considered cancelled between each other in the partition function\footnote{For a more detailed explanation see \cite{Kunz:2025}.}. Therefore, all propagators in the amplitudes, including the ones of the auxiliary fields, are simply changed to Szeg\"{o} kernels.\\
  
  One important change for the one-loop level is that every gauge symmetry has one zero mode for both the ghost and anti-ghost. Although accompanied by picture changing operators (PCOs), these pairs of ghost and anti-ghost cannot be used to change fixed vertex operators to integrated ones. However, they become available during non-separating degeneration in section \ref{NonSeparating}.\\
  
  It follows that the one-loop amplitude looks like the tree amplitude \eqref{treeAmpl} with SL$(2,\mathbb{C})_{\sigma}$ replaced with GL$(1,\mathbb{C})_{\sigma}$, all propagators $1/{\sigma_{rl}}$ with Szeg\"{o} kernels $S_\alpha(\sigma_r, \sigma_l; \tau)\!/\!(\sqrt{\ud \sigma_r}\sqrt{\ud \sigma_l})$, with an additional set of PCOs without anti-ghosts, with an additional integration over the torus modulus $\tau$ and partition functions, and finally with integration over zero modes for the odd spin structure, divided by the volume of the little group because of invariance under transformations of this group.\\
  
  The one-loop amplitude for even and odd spin structures becomes \cite{Kunz:2025}
  \begin{align}
  \label{1loopAmpl}
  \mathcal{A}_{\text{even},n}^{\text{1-loop}} \!=\! \!\sum_{\alpha} \! \int &\ud \mu^{\text{pol}}_{n|\text{even}} \Bigl(\frac{\theta[\alpha](0)}{\eta(\tau)^3}\Bigr)^{\!8} \!q_\alpha^2 \Biggl(\Bigl \langle  \!\bar{\delta}_x \, j_z \tilde{j}_{\tilde{z}} \!\Bigr\rangle_{\!\!\alpha} \text{det}^\prime H \;\text{det}^\prime G + \cdots \Biggr),\;q_\alpha \!=\! \begin{cases}
  e^{\pi i \tau}, \alpha = 3,4,\\
  1,\quad\, \alpha = 2
  \end{cases}\nonumber\\
  \ud \mu^{\text{pol}}_{n|\text{even}} &= \! \ud \tau \prod_{l=1}^n \frac{\ud \sigma_l \ud^2 u_l \ud^2 v_l}{\text{vol}(\text{GL}(1,\mathbb{C})_{\sigma} \!\times\! \text{SL}(2,\mathbb{C}))_u} \begin{cases}
  \ud \nu^{\text{pol}}_{n|\text{even}} \\
  \ud \nu^{\text{pol}|\mathcal{N}}_{n|\text{even}}
  \end{cases}\nonumber,\\
  &\ud \nu^{\text{pol}}_{n|\text{even}} =  \prod_{r = 1}^n \!\bar{\delta}(\!\braket{v_r \epsilon_r} \!-\!1\!) \bar{\delta}^{4}(\!\braket{u_r \lambda^A} \!-\! \braket{v_r \kappa_r^A}\!) \;e^{F^{\text{even}}_{\mathcal N}} ,\nonumber\\
  &\ud \nu^{\text{pol}|\mathcal{N}}_{n|\text{even}} =  \prod_{r = 1}^n \!\bar{\delta}(\!\braket{v_r \epsilon_r} \!-\!1\!) \bar{\delta}^{4}(\!\braket{u_r \lambda^A} \!-\! \braket{v_r \kappa_r^A}\!) \bar{\delta}^{\mathcal{N}}\!(\!\braket{u_r \eta^{\mathcal{I}}} \!-\! \braket{v_r \zeta_r^{\mathcal{I}}}\!) ,\nonumber\\
  F_{\mathcal{N}}^{\text{even}} =& \sum_{i < j} \braket{u_i u_j} q_{i\mathcal{I}} q_j^{\mathcal{I}}\, \frac{S_{\alpha}(\sigma_i, \sigma_j; \tau)}{\sqrt{\ud \sigma_i}\sqrt{\ud \sigma_j}}  - \frac{1}{2}\sum_{i = 1}^n \braket{\xi_i v_i} q_i^2,\nonumber\\
  \mathcal{A}_{\text{odd},n}^{\text{1-loop}} \!=\delta^4\!\Bigl(\sum_{i=1}^n&\braket{\kappa_i^{\alpha} \tilde{\kappa}_i^{\dot\alpha}}\!\Bigr) \delta^{4\mathcal{N}}\!\Bigl(\sum_{i=1}^n \!\braket{\kappa_{iA} \zeta_{i\mathcal{I}}}\!\Bigr)\!\! \int\!\! \ud \mu^{\text{pol}}_{n|\text{odd}} \Biggl(\Bigl \langle  \!\bar{\delta}_x \, j_z \tilde{j}_{\tilde{z}} \!\Bigr\rangle\text{det}^\prime H \;\text{det}^\prime G + \cdots \Biggr), \\
  \ud \mu^{\text{pol}}_{n|\text{odd}} &=\!\frac{\ud^8 \!\lambda_{0A}^a \; \ud^{2\mathcal{N}} \!\eta_{0\mathcal{I}}^a \; \ud \tau}{\text{vol}(\text{SL}(2,\mathbb{C}) \times \mathbb{C})_{\lambda_0}} \prod_{l=1}^n \frac{\ud \sigma_l \ud^2 u_l \ud^2 v_l}{\text{vol}(\text{GL}(1,\mathbb{C})_{\sigma} \!\times\! \text{SL}(2,\mathbb{C}))_u} \begin{cases}
  \ud \nu^{\text{pol}}_{n|\text{odd}} \\
  \ud \nu^{\text{pol}|\mathcal{N}}_{n|\text{odd}}
  \end{cases}\nonumber,\\
  &\ud \nu^{\text{pol}}_{n|\text{odd}} =  \prod_{r = 1}^n \!\bar{\delta}(\!\braket{v_r \epsilon_r} \!-\!1\!) \bar{\delta}^{4}(\!\braket{u_r \lambda^A} \!-\! \braket{v_r \kappa_r^A}\!) \;e^{\frac{1}{2}\!\braket{\eta_{0\mathcal{I}} \eta_0^{\mathcal{I}}} + F^{\text{odd}}_{\mathcal N}} ,\nonumber\\
  &\ud \nu^{\text{pol}|\mathcal{N}}_{n|\text{odd}} = \prod_{r = 1}^n \!\bar{\delta}(\!\braket{v_r \epsilon_r} \!-\!1\!) \bar{\delta}^{4}(\!\braket{u_r \lambda^A} \!-\! \braket{v_r \kappa_r^A}\!) \bar{\delta}^{\mathcal{N}}\!(\!\braket{u_r \eta^{\mathcal{I}}} \!-\! \braket{v_r \zeta_r^{\mathcal{I}}}\!) ,\nonumber\\
  F_{\mathcal{N}}^{\text{odd}} =& \sum_{i < j} \braket{u_i u_j} q_{i\mathcal{I}} q_j^{\mathcal{I}}\, \frac{S_1(\sigma_i, \sigma_j; \tau)}{\sqrt{\ud \sigma_i}\sqrt{\ud \sigma_j}} + \sum_{i = 1}^n \braket{\eta_{0\mathcal{I}} u_i} q_i^{\mathcal{I}} - \frac{1}{2}\sum_{i = 1}^n \braket{\xi_i v_i} q_i^2,\nonumber\\
  \bar{\delta}_x \!=& \bar{\delta}(\!\braket{\lambda^A(x) \lambda_A(x)}\!)\bar{\delta}(\!\braket{\lambda^A(\tilde{x}) \lambda_A(\tilde{x})}\!+\!\braket{\eta^{\mathcal{I}}(\tilde{x}) \eta_{\mathcal{I}}(\tilde{x})}\!),\nonumber\\
  j_z =& \braket{\lambda^A(z) \rho_{1A}(z)} \int \!\frac{\ud^2 u_1 \ud^2 u_2 \ud^2 u_3e^{-\sum_{k=1}^3 ||u_{k}||^2}}{\braket{u_1 u_2}\braket{u_2 u_3}\braket{u_3 u_1}}\prod_{k=1}^3 (\braket{u_k \lambda^A(z_k)}\braket{u_k \rho_{1A}(z_k)})\,,\nonumber\\
  \tilde{j}_{\tilde{z}} =& (\braket{\lambda^A(\tilde{z}) \rho_{2A}(\tilde{z})}\!+\!\braket{\eta^{\mathcal{I}}(\tilde{z}) \tau_{\mathcal{I}}(\tilde{z})})\; \int \!\!\frac{\ud^2 \tilde{u}_1 \ud^2 \tilde{u}_2 \ud^2 \tilde{u}_3}{\braket{\tilde{u}_1 \tilde{u}_2}\braket{\tilde{u}_2 \tilde{u}_3}\braket{\tilde{u}_3\tilde{u}_1}}e^{-\sum_{k=1}^3 ||\tilde{u}_{k}||^2}\nonumber\\
  & \prod_{k=1}^3 (\braket{\tilde{u}_k \lambda^{A}(\tilde{z}_k)}\braket{\tilde{u}_k \rho_{kA}(\tilde{z}_k)} \!+\! \braket{\tilde{u}_k \eta^{\mathcal{I}}(\tilde{z}_k)} \braket{\tilde{u}_k \tau_{\mathcal{I}}(\tilde{z}_k)})\,,\nonumber\\
  H_{ij} =& \epsilon_{iA} \epsilon_j^A \frac{S_{\alpha}(\sigma_i, \sigma_j; \tau)}{\sqrt{\ud \sigma_i}\sqrt{\ud \sigma_j}}, \qquad\qquad\quad\, u_i^a H_{ii} = - \lambda_A^a \epsilon_i^A,\nonumber\\
  G_{ij} =& (\epsilon_{iA} \epsilon_j^A + q_{i\mathcal{I}} q_j^\mathcal{I}) \frac{S_{\alpha}(\sigma_i, \sigma_j; \tau)}{\sqrt{\ud \sigma_i}\sqrt{\ud \sigma_j}}, \quad u_i^a G_{ii} = - \! \lambda_A^a \epsilon_i^A - \eta^a_\mathcal{I} q_i^\mathcal{I}.\nonumber
  \end{align}
    Here $\lambda^a_A$ and $\eta^a_{\mathcal{I}}$ are to be inserted from \eqref{scattEq1L}. For convenience, the PCOs \,$\!\bar{\delta}_x \, j_z \tilde{j}_{\tilde{z}}$ have been listed before any contractions among the auxiliary spinors; they have been made invariant under little group transformations, their arguments are arbitrary, and the ellipsis stands for terms with contractions between auxiliary spinors and integrated vertex operators \cite{Kunz:2025}. $\theta[\alpha](0)$ is the theta constant for the even spin structure with characteristic $\alpha$ and $\eta(\tau)$ is the Dedekind eta function. For the amplitude $\mathcal{A}_{\text{odd},n}^{\text{1-loop}}$ on the odd spin structure momentum and supercharge conservation do not follow automatically from the scattering equations. Momentum conservation is expressed explicitly with a delta function, coming from the integration over the zero modes $\mu^{aA}_0$, and similarly a delta function for super charges arising from integration over the zero modes $\tilde{\eta}_a^\mathcal{I}$ \cite{Kunz:2025}. The measures are expressed in both the LG representation and the R-symmetry preserving representation, by applying the Fourier transformation that transitions from LG to R-symmetry similar to \eqref{Fourier} for tree amplitudes. $F_{\mathcal{N}}^{\text{odd}}$ requires some further explanation. The Fourier transformation \eqref{Fourier} only provides half of the second term $\sum_{i = 1}^n \braket{u_i \eta_{0\mathcal{I}}} q_i^{\mathcal{I}}$. The other half actually can originate from Fourier transforming the zero mode $\eta_0$ itself, according to:
  \begin{equation*}
  \int \ud^{2N} \!\eta_{0a}^I  \;\; 2^{-2N} \!\int \ud^{2N} \!\tilde{\eta}_{0a}^I  e^{\frac{1}{2}\!\braket{\tilde{\eta}_0^{\mathcal{I}} \eta_{0\mathcal{I}}} + \frac{1}{2}\sum_{i = 1}^n \!\braket{\tilde{\eta}_0^{\mathcal{I}} u_i} q_{i\mathcal{I}}} = 1\,.
  \end{equation*}
     An exponential factor similar to exp($\frac{1}{2}\!\braket{\eta_{0\mathcal{I}} \eta_0^{\mathcal{I}}}$) appears in \cite{Adamo:2015, Albonico:2023}.\\
    
  The amplitude $\mathcal{A}_n^{\text{1-loop}} $ is modular invariant \cite{Kunz:2025} such that the integration over the modulus can be restricted to a fundamental domain. The delta functions in $\bar{\delta}_x$ serve to localize the modulus.\\
 
 \subsection{Non-separating Degeneration}
\label{NonSeparating}
This section checks whether the one-loop amplitude \eqref{1loopAmpl} factorizes properly for a non-separating degeneration of the Riemann surface. Pinching a non-separating cycle occurs in the limit $q = e^{\pi i \tau} \rightarrow 0$. For the odd spin structure, one obstruction to dealing with this limit by simply taking the residue at 0 of the integrand as a function in $q$ is that the integrand is not an analytic function in $q$ because the torus Szeg\"{o} kernel $S_1(\sigma_1, \sigma_2; \tau)$ contains an Im($\tau$)-dependent term:
  \begin{equation}
  \label{S1}
  \frac{S_1(\sigma_1, \sigma_2; \tau)}{\sqrt{\ud\sigma_i \ud\sigma_j}} = \left(\frac{\theta_1^{'}\!(\sigma_{12}, \tau)}{\theta_1(\sigma_{12}, \tau)} + 4 \pi \frac{\mathrm{Im}(\sigma_{12})}{\mathrm{Im}(\tau)} \right)\,.
  \end{equation}
The behavior of $S_1$ can be examined by modeling the torus as a Riemann sphere with a handle whose ends are located at two reference points $\sigma_{+}$ and $\sigma_{-}$, i.e. these two points are added as additional punctures, modeled in terms of local coordinates by
 \begin{equation*}
 (\sigma - \sigma_{+})(\sigma - \sigma_{-}) = q.
 \end{equation*}
 Then, because according to \cite{Tuite:2010} all Szeg\"o kernels become holomorphic in $q$ on the degenerated Riemann surface for small enough $q$, the integrand in \eqref{1loopAmpl} becomes a manifestly meromorphic function in $q$. In general, even for higher loop amplitudes, Szeg\"{o} kernels on odd spin structures are replaced by meromorphic functions in all arguments that correspond to the first term on the right hand side of \eqref{S1}\cite{Kunz:2025}.\\
 
  The asymptotic behavior of the Szeg\"o kernel on all spin structures is \cite{Casali:2014, Geyer:2016}
 \begin{equation*}
 \begin{aligned}
 &S_{\alpha}(\sigma_i, \sigma_j; \tau) \rightarrow &\frac{\sqrt{\ud z_i\ud z_j}}{z_{ij}}\;+\, O(q),
 \end{aligned}
 \label{asymptotic}
 \end{equation*}
 in terms of the coordinates $z = e^{2 \pi i (\sigma \!-\! \frac{\tau}{2})}$ on the Riemann sphere i.e. $S_{\alpha}$ becomes just the genus 0 propagator in this limit.\\
 
 Considering first the case of the odd spin structure, on the Riemann sphere the torus zero modes $\lambda_0$ and $\eta_0$ are lifted up to fields $\lambda_0/z$ and $\eta_0/z$. To proceed, the points $z_{+}$ and $z_{-}$ are chosen temporarily as 0 and $\infty$. Then the right hand side of the scattering equations \eqref{scattEq1L} for $\lambda$ can be rewritten as, in abuse of notation changing back from $z$ to $\sigma$,
 \begin{equation*}
  \lambda^{aA}(\sigma) = \frac{\lambda^{aA}_0}{\sigma} + \sum_l \frac{u^a_l \epsilon^A_l}{\sigma - \sigma_l} + O(q) = \frac{\braket{u^a_{+} \epsilon^A_{+} + u^a_{-}\epsilon^A_{-}}}{\sigma} + \sum_l \frac{u^a_l \epsilon^A_l}{\sigma - \sigma_l} + O(q),
  \end{equation*}
  where $\lambda^{aA}_0$ has been expanded into a basis of little spinors $u^a_{\pm}$ with $\braket{u_{+} u_{-}} \!\ne\! 0$. Now reverting to general $\sigma_{+}$ and $\sigma_{-}$, designating $u^a_{\pm} \epsilon^A_{\pm}$ to belong to $\sigma_{\pm}$ by requiring that $\braket{u_{\pm} \lambda^A(\sigma_{\pm})} \!<\! \infty$, defining a spinor basis $\epsilon^a_{\pm}$ with $\braket{\epsilon_{+} \epsilon_{-}} \!=\! 1$ and momenta $\kappa^a_{\pm A}$ with $\kappa^1_{+A} = \kappa^1_{-A}, \kappa^2_{+A} = - \kappa^2_{-A}$ such that $\epsilon^A_{\pm} \!=\! \braket{\epsilon_{\pm} \kappa^A_{\pm}}$, and finally viewing $\sigma_{\pm}$ as additional punctures $\sigma_{n\!+\!1}$ and $\sigma_{n\!+\!2}$ one gets
  \begin{equation}
  \label{lambda0}
  \lambda^{aA}(\sigma) = \sum_{l=1}^{n+2} \frac{u^a_l \epsilon^A_l}{\sigma - \sigma_l} + O(q) .
  \end{equation}
  A similar procedure for $\eta_0$ gives
  \begin{equation}
  \label{eta0}
  \eta^{a\mathcal{I}}(\sigma) \!=\! \sum_{l=1}^{n+2} \frac{u^a_l q^{\mathcal{I}}_l}{\sigma - \sigma_l} \!+\! O(q) \;\;\text{with}\;q^{\mathcal{I}}_{n\!+\!\genfrac{}{}{0pt}{}{1}{2}} \!=\! \braket{\epsilon_{n\!+\!\genfrac{}{}{0pt}{}{1}{2}} \zeta^{\mathcal{I}}_{n\!+\!\genfrac{}{}{0pt}{}{1}{2}}}, \zeta^1_{n\!+\!1 \mathcal{I}} \!=\! \zeta^1_{n\!+\!2 \mathcal{I}}, \zeta^2_{n\!+\!1 \mathcal{I}} \!=\! -\zeta^2_{n\!+\!2 \mathcal{I}}.
  \end{equation}
  This kind of degeneration of the zero modes into differential forms interpolating between the two endpoints of the handle is a general occurrence that takes place for higher loop amplitudes as well \cite{Kunz:2025}. When defining a total momentum operator $P^{AB}$ as
  \begin{equation*}
  P^{AB}(\sigma) = \braket{\lambda^A(\sigma) \lambda^B(\sigma)},
  \end{equation*}
  it follows $P^2(\sigma) \!=\! 0$ and from the presence of $\bar{\delta}(\!\braket{\lambda^A(\sigma) \lambda_A(\sigma})$ in the amplitude also that $P^A_A(\sigma) \!=\! 0$. From the latter and the freedom in the choice of $u_{\pm}$, polarized scattering equations for the zero modes and, therefore, for all $n\!+\!2$ particles are obtained \cite{Kunz:2025}. Furthermore, the corresponding polarized scattering equations for $\eta_{\mathcal{I}}^a(\sigma)$ are consistent with the condition $\bar{\delta}(\!\braket{\eta^\mathcal{I}(\sigma) \eta_\mathcal{I}(\sigma})$.\\
  
  What happens to $F_{\mathcal{N}}^{\text{odd}}$? Inserting
  \begin{equation}
  \label{eta0RiemSph}
  \eta_0(\sigma) = u_{n\!+\!1} q_{n\!+\!1} /(\sigma - \sigma_{n\!+\!1}) + u_{n\!+\!2} q_{n\!+\!2} /(\sigma - \sigma_{n\!+\!2})
   \end{equation}
   from \eqref{eta0} into $\braket{\eta_{0\mathcal{I}} \eta_0^{\mathcal{I}}}$ and $F_{\mathcal{N}}^{\text{odd}}$ in \eqref{1loopAmpl} gives $F_{\mathcal{N}}$ of the tree amplitude \eqref{treeAmpl1}:
  \begin{align*}
  &\frac{1}{2}\!\braket{\eta_{0\mathcal{I}} \eta_0^{\mathcal{I}}} \xrightarrow{\eqref{eta0RiemSph}} \frac{1}{2}(q_{n\!+\!1}^{\mathcal{I}}\!\braket{u_{n\!+\!1} \eta_{0 \mathcal{I}}(\sigma_{n\!+\!1})} + q_{n\!+\!2}^{\mathcal{I}}\!\braket{u_{n\!+\!2} \eta_{0 \mathcal{I}}(\sigma_{n\!+\!2})}) = \frac{\braket{u_{n\!+\!1} u_{n\!+\!2} }}{\sigma_{n\!+\!1 n\!+\!2}}q_{n\!+\!1 \mathcal{I}} q_{n\!+\!2}^{\mathcal{I}},\\
  &\frac{1}{2}\!\braket{\eta_{0\mathcal{I}} \eta_0^{\mathcal{I}}} \!+\! F_{\mathcal{N}}^{\text{odd}} =\!\! \sum_{1 \leq i < j \leq n\!+\!2} \!\!\frac{\braket{u_i u_j}}{\sigma_{ij}}q_{i\mathcal{I}} q_j^{\mathcal{I}} - \frac{1}{2}\sum_{i = 1}^{n\!+\!2} \braket{\xi_i v_i} q_i^2 = F_{\mathcal{N}},
  \end{align*}
  where it was used that $\sum_{i = n\!+\!1}^{n\!+\!2} \braket{\xi_i v_i} q_i^2 \!=\! 0$ because of $\braket{\xi v}\!\epsilon^a \!=\! v^a \!-\! \xi^a$ and the special relationship between values for $\zeta_{n\!+\!1}$ and $\zeta_{n\!+\!2}$ in \eqref{eta0}.\\
  
  Now the measure needs to be modified. This is done by setting
  \begin{align*}
  \delta^4 \!\!\left(\sum_{i=1}^n \braket{\kappa_i^{\alpha} \tilde{\kappa}_i^{\dot\alpha}}\right) &\delta^{4\mathcal{N}}\!\Bigl(\sum_{i=1}^n \!\braket{\kappa_{iA} \zeta_{i\mathcal{I}}}\!\Bigr)\frac{\ud^8 \!\lambda_{0A}^a \,\ud^{2\mathcal{N}} \!\eta_{0\mathcal{I}}^a}{\text{vol}(\text{GL}(1, \mathbb{C})_{\sigma})} \frac{\ud q}{q}
  \!=\! \int \!\!\frac{\prod_{l= n\!+\!1}^{n\!+\!2} \!\ud \sigma_l \,\ud^2 u_l \,\ud^2 v_l }{\text{vol}(\text{SL}(2, \mathbb{C})_{\sigma} \!\times\! \text{SL}(2, \mathbb{C})_u)}\\
  &\frac{\prod_{l= n\!+\!1}^{n\!+\!2} \ud^4 \epsilon_l^A \,\ud^{\mathcal{N}} \!q_l^{\mathcal{I}}}{\braket{u_{n\!+\!1} u_{n\!+\!2}}^4} \prod_{r = n\!+\!1}^{n\!+\!2} \!\bar{\delta}(\!\braket{v_r \epsilon_r} \!-\!1) \bar{\delta}^{4}(\!\braket{u_r \lambda^A} \!-\! \braket{v_r \kappa_r^A}\!)\bar{\delta}^{\mathcal{N}}\!(\!\braket{u_r \eta^{\mathcal{I}}} \!-\! \braket{v_r \zeta_r^{\mathcal{I}}}\!).
  \end{align*}
  That the left hand side is equal to the right hand side can be seen as follows: on the tree level two more $\sigma_l$ values can be fixed (as was done with $\sigma_{n\!+\!1}$ and $\sigma_{n\!+\!2}$) and three of the $u^a_l$ variables (e.g. by choosing $u^a_{n\!+\!1} \!=\! (1, 0),  u^a_{n\!+\!2} = (u_{n\!+\!2}, 1)$), fixing 5 of the 10 bosonic delta functions on the right. On the left, because momentum conservation follows from the scattering equations for the $n\!+\!2$ particles and division by vol(GL$(1, \mathbb{C})_{\sigma}$) is equivalent to a delta function, the remaining 5 delta functions on both sides balance each other out. Finally $\braket{u_{n\!+\!1} u_{n\!+\!2}}^{4-\mathcal{N}}$ = $\braket{u_{n\!+\!1} u_{n\!+\!2}}^{-4}$ is the Jacobian going from $\ud^8 \!\lambda_{0} \,\ud^{2\mathcal{N}} \!\eta_{0}$ to $\prod_l \ud^4 \epsilon_l \,\ud^{\mathcal{N}} \!q_l$. The fact that $\mathcal{N} = 8$ is crucial here.\\
  
  The right hand side of the equation shows that the non-separating degeneration of the zero modes on a pinched cycle is equivalent to the insertion of fixed vertex operators at the punctures $\sigma_{+}$ and $\sigma_{-}$ representing a complete set of physical states:
 \begin{align*}
\label{Completeness}
\int \!\frac{\prod_{l= n\!+\!1}^{n\!+\!2} \ud^4 \epsilon_l^A \,\ud^{\mathcal{N}} \!q_l^{\mathcal{I}}}{\text{vol}(\text{SL}(2, \mathbb{C})\times \mathbb{C})_{\epsilon}} \;&\!\!\prod_{l= n\!+\!1}^{n\!+\!2} \!\!\ud^2 u_l\;\ud^2\!v_l\;c(\sigma_{n+1})c(\sigma_{n+2})c\mathcal{W}(u) \bar{\delta}(\!\braket{v_l \epsilon_l} \!-\!1) \bar{\delta}^{4}(\!\braket{u_l \lambda_A(\!\sigma_l\!)} \!-\! \braket{v_l \kappa_{lA}}) \nonumber\\
&\bar{\delta}^{\mathcal{N}}\!(\!\braket{u_l \eta_{\mathcal{I}}(\!\sigma_l\!)} \!-\! \braket{v_l \zeta_{l\mathcal{I}}}) \; e^{\braket{u_l \mu^A(\!\sigma_l\!)} \epsilon_{lA} + \braket{u_l \tilde{\eta}^{\mathcal{I}}(\!\sigma_l\!)} q_{l\mathcal{I}}},
\end{align*}
where two additional $c$ ghosts were added to make up for vol(SL(2, $\mathbb{C}$)$_{\sigma}$).
Such an insertion actually needs to be done for the degeneration on even spin structures\footnote{More details and generalizations to higher loop amplitudes are given in \cite{Kunz:2025}.}. The result is that for both types of spin structures the outcome is a tree amplitude for $n\!+\!2$ particles, except for a few obstructions: the PCOs from the one-loop level are still around and do not belong on the tree level, then there are now two sets of fixed vertex operators, one inherited from the one-loop level and one from the inserted set of physical states, and finally there is a remnant $q_\alpha^2 \!(\theta[\alpha]/\eta)^8$ of the partition functions for even spin structures (the remnant is 1 for the odd spin structure). To deal with this, one has to note that, in the process of inserting the complete set of physical states, ghost fields are inserted as well, making up the exact number of ghost zero modes necessary for the tree level such that on the degenerated Riemann surface the partition functions for the ghosts can be considered dissolved by releasing the ghost and anti-ghost zero modes to make them available for changing fixed vertex operators of one set to being integrated, consuming all the PCOs. Because of $q_\alpha^2 \!(\theta[\alpha]/\eta^3)^8 \sim 1 (q \!\rightarrow\! 0)$ for all $\alpha$'s, all four types of spin structures give the same finite result.\\
  
  In summary, the one-loop amplitude \eqref{1loopAmpl} factorizes correctly into a tree amplitude \eqref{treeAmpl1} where two additional punctures in the forward direction are connected through a complete set of intermediate physical states:
  \begin{align*}
  \mathcal{A}_{\text{odd},n}^{\text{1-loop}} = \frac{1}{3} \mathcal{A}_{\text{even},n}^{\text{1-loop}} = &\frac{1}{4}\int \frac{\prod_{l= n\!+\!1}^{n\!+\!2} \ud^4 \!\epsilon_l^A \,\ud^{\mathcal{N}} \!q_l^{\mathcal{I}}}{\text{vol}(\text{SL}(2,\mathbb{C}) \times \mathbb{C})_{\lambda_0}} \;\mathcal{A}_{n+2}^{\text{tree}}, \qquad\quad\braket{\epsilon_{n+1} \epsilon_{n+2}} = 1,\\
  &\epsilon^A_{n\!+\!\genfrac{}{}{0pt}{}{1}{2}} \!=\! \braket{\epsilon_{n\!+\!\genfrac{}{}{0pt}{}{1}{2}} \kappa^A_{n\!+\!\genfrac{}{}{0pt}{}{1}{2}}}, \kappa^1_{n\!+\!1 A} \!=\! \kappa^1_{n\!+\!2 A}, \kappa^2_{n\!+\!1 A} \!=\! -\kappa^2_{n\!+\!2 A},\\
  &q^{\mathcal{I}}_{n\!+\!\genfrac{}{}{0pt}{}{1}{2}} \!=\! \braket{\epsilon_{n\!+\!\genfrac{}{}{0pt}{}{1}{2}} \zeta^{\mathcal{I}}_{n\!+\!\genfrac{}{}{0pt}{}{1}{2}}}, \,\zeta^1_{n\!+\!1 \mathcal{I}} =\! \zeta^1_{n\!+\!2 \mathcal{I}}, \;\zeta^2_{n\!+\!1 \mathcal{I}} =\! -\zeta^2_{n\!+\!2 \mathcal{I}}.
  \end{align*}
  \ \\

 \subsection{Separating Degeneration}                                    
 \label{Separating}
 Pinching a separating cycle involves the degeneration of the Riemann surface $\Sigma$ of genus $g \!=\! g_L \!+\! g_R$ into a surface $\Sigma_L$ of genus $g_L$ and a surface $\Sigma_R$ of genus $g_R$ and can again be modeled in terms of local coordinates by
 \begin{equation*}
 (\sigma - \sigma_L)(\sigma - \sigma_R) = q\,,
 \label{sepDeg}
 \end{equation*}
 where $\sigma_L$ is an extra puncture on $\Sigma_L$ and $\sigma_R$ one on $\Sigma_R$ with their neighborhoods sewed together and both points identified in the pinching limit $q \rightarrow 0$.\\
 
Proper factorization requires to show that during such a separating degeneration an $n$-particle amplitude on $\Sigma$ can be represented as an $n_L\!+\!1$-particle amplitude on $\Sigma_L$ and an $n_R\!+\!1$-particle amplitude on $\Sigma_R$ with $n \!=\! n_L \!+\! n_R$ and additional punctures $\sigma_L$ and $\sigma_R$ which are connected through insertion of a complete set of physical states.\\
 
 Factorization for separate degeneration of a one-loop amplitude into a tree and a one-loop amplitude can be simply demonstrated by first factorizing the one-loop amplitude into a tree amplitude using non-separating factorization, applying factorization of tree amplitudes from subsection \ref{TreeFactorization} with the two additional punctures from the loop amplitude necessarily ending up on the same side, and then reversing the non-separating degeneration for that side, possibly with a reduced number of punctures.\\
 
 \subsection{Higher Loops and Cosmological Constant}
 \label{hLoops}
 Looking first at partition functions at two or more loops, there is a strictly positive number of zero modes of anti-ghosts and, therefore, at the minimum one picture-changing operator for every supersymmetric gauge symmetry, resulting in zero on even spin structures because of no contractions possible like at the one-loop level. Because the partition function on odd spin structures is always zero due to the presence of a Dirac zero mode, the cosmological constant vanishes at every order of perturbation theory. Of course, non-perturbative effects like IR renormalization for massless particles cannot be excluded and might give the cosmological constant a non-vanishing value.\\
 
 Multi-loop amplitudes with a non-zero number of vertex operators will again generally develop a single zero mode for the $\lambda^a_A$ and $\eta^a_{\mathcal{N}}$ fields on odd spin structures. Assuming that non-separating degeneration follows the same pattern as on the one-loop level, amplitudes can be fully reduced to the tree level with $n \!+\! 2g$ external particles of which $g$ particle $\!-\!$ anti-particle pairs are integrated over in the forward direction\footnote{This is actually a correct assumption \cite{Kunz:2025}.}.\\
 
 It is interesting to note that many of the results in this article also apply to the original massive ambitwistor model of \cite{Albonico:2020, Albonico:2022, Albonico:2023}. This includes zero cosmological constant and proper factorization of one-loop amplitudes on even spin structures.\\

 \section{Compton Scattering Using Massive Twistor Strings}
  \label{massiveCompton}
  In this section it is important to note that in the following the usual convention is used that in amplitudes all legs are taken to be outgoing, i.e. incoming particles will have their momenta and helicities reversed.\\
         
  The 4-point amplitude for the original massive ambitwistor model of \cite{Albonico:2020, Albonico:2022, Albonico:2023} is calculated to be \cite{Albonico:2020}
 \begin{equation}
 \label{massComptonAmpl}
 \begin{split}
  &M_4 = \frac{\langle 1 2 3 4 \rangle^4} {(2p_1 \!\cdot\! p_2 ) (2p_2 \!\cdot\! p_3) (2p_2 \!\cdot\! p_4)} e^{F_\mathcal{N}}, \\
  &\langle 1 2 3 4 \rangle = \varepsilon^{ABCD} e_{1A} e_{2B} e_{3C} e_{4D},\\
  &F_{\mathcal{N}} = \frac{1}{2} \sum_{i,j = 1}^4 \frac{\braket{u_i u_j}}{\sigma_{ij}} q_{i\,\mathcal{I}} q_j^{\mathcal{I}} - \frac{1}{2} \sum_{i=1}^4 \braket{\xi_i v_i} q_i^2 .
  \end{split}
 \end{equation}
 More details can be found in \cite{Albonico:2020}\footnote{In \cite{Albonico:2020} the amplitude is given for 6 dimensions, with $\langle 1 2 3 4 \rangle^4$ replaced by $\langle 1 2 3 4 \rangle^2[1 2 3 4]^2$, and $F_\mathcal{N}$ with  $F_N + \tilde{F}_N$, but the contributions become equal via dimensional reduction.}.\\
 
 The amplitude \eqref{massComptonAmpl} is valid for 4 massive particles up to spin 2. It is now specialized to the case of Compton scattering of a massless particle with momenta $p_2 \!\rightarrow\! k_2, p_3 \!\rightarrow\! k_3$ against a heavy massive particle with mass $M$ much larger than the energy $\hbar \omega$ of the massless particle. Switching from the polarized formalism to a standard little-group based amplitude, it can readily be seen that the amplitude becomes either
 \begin{equation}
 \label{oppHelAmpl}
 M_4^{+-} = \frac{(\langle \mathbf{4} 3 \rangle \, [\mathbf{1} 2] - \langle \mathbf{1} 3 \rangle \, [\mathbf{4} 2])^4}{(2p_1 \!\cdot\! k_2 ) (2k_2 \!\cdot\! k_3) (2k_2 \!\cdot\! p_4)} e^{F_\mathcal{N}}
 \end{equation}
 or 
 \begin{equation}
 \label{sameHelAmpl}
 M_4^{++} = \frac{\langle \mathbf{1} \mathbf{4} \rangle^4 \, [ 2 3]^4 }{(2p_1 \!\cdot\! k_2 ) (2k_2 \!\cdot\! k_3) (2k_2 \!\cdot\! p_4)} e^{F_\mathcal{N}},
 \end{equation}
 depending on whether the incoming massless particle with negative helicity preserves or flips the helicity, respectively\footnote{Remember from the note at the beginning of the section that because of the convention to regard all particles outgoing, conserving/flipping the helicity means that the incoming and outgoing massless particles have the opposite/same helicity.}. In order to simplify formulas, the same notation as in \cite{Arkani-Hamed:2017jhn} is used by bolding the massive spinor helicity variables and suppressing the little group indices. \eqref{oppHelAmpl} matches the amplitude (5.35) in \cite{Arkani-Hamed:2017jhn}, up to the supersymmetric exponential factor.\\
 
 The supersymmetric factor is useful when trying to switch from spin 2 to lower spin particles, by taking derivatives with respect to the supermomenta $q_\mathcal{I}$ and setting them to zero. It was shown in \cite{Albonico:2020} that in order to lower the spin of the massless particle it is required to multiply the amplitude with a factor of
 
 \begin{equation*}
 \frac{\braket{u_2 u_3}}{\sigma_{23}} = -\frac{\langle k_2 1 4 \rangle}{\langle 1 2 3 4 \rangle}
 \end{equation*}
 
 for every spin $\frac{1}{2}$ downward change and for the massive particle with a factor of
 
 \begin{equation*}
 \frac{\braket{u_1 u_4}}{\sigma_{14}} = -\frac{\langle p_1 2 3 \rangle}{\langle 1 2 3 4 \rangle}.
 \end{equation*}
 
 Because $\langle k_2 1 4 \rangle = 0$ when the massless particle flips its helicity if follows that the amplitude \eqref{sameHelAmpl} vanishes for massless particles with spin$< \!2$. Therefore, in this model only a graviton can scatter with reversing its helicity \cite{Bautista:2021wfy}. \\
 
 Concerning the amplitude that conserves helicity, when disregarding the difference between $p_1 \!\cdot\! k_2$ and $p_1 \!\cdot\! k_3$ (valid in the classical limit $\hbar \omega / M \rightarrow 0$) in equation (5.26) of \cite{Albonico:2020} when applied to 4 dimensions:
 \begin{equation}
 \label{auxHel}
  \langle k_2 1 4 \rangle \langle p_1 2 3 \rangle = -\!2 p_1 \!\cdot\! k_3 \; e_{1A} e_2^A e_{4B} e_3^B  -\!2 p_1 \!\cdot\! k_2 \; e_{1A} e_3^A e_{4B} e_2^B \rightarrow -\!2 p_1 \!\cdot\! k_3 \langle 1 2 3 4 \rangle ,
 \end{equation}
 and switching to the little-group helicity-based formalism, the amplitude \eqref{oppHelAmpl} becomes
 \begin{equation}
 \label{ComptAmpl}
 M_4^{+-} = \frac{\langle 3 | p_1 | 2]^{2h} (2p_1 \!\cdot\! k_3)^{2(2-h)}}{(2p_1 \!\cdot\! k_2 ) (2k_2 \!\cdot\! k_3) (2k_2 \!\cdot\! p_4)} \left(\frac{\langle \mathbf{4} 3 \rangle \, [\mathbf{1} 2] - \langle \mathbf{1} 3 \rangle \, [\mathbf{4} 2]}{\langle 3 | p_1 | 2]}\right)^{2s},
 \end{equation}
where $s$ is the spin of the massive particle and $h$ the spin of the massless particle.\\

What is different with the model of this article? Assume first that the fermionic contribution in det$^{\prime} G$ in \eqref{Matrices} is for a particle of spin 3/2 ('gravitino') or spin 1 ('gluon') still preserving helicity. This represents a massive EYM amplitude as mentioned in subsection \ref{consChk} with a 2-particle PT factor for the 'gluon' and an open trace for the 'gravitino':
\begin{equation*}
\label{gluonGrav}
\text{det}^\prime G \rightarrow \begin{cases}
       \left(\!\frac{q_{2\mathcal{I}} q_3^{\mathcal{I}}}{\sigma_{23}}\right)^{\!2} \text{det}H^{[23]}_{[32]} = \text{det}^{\prime} H \!\left(\!\frac{\braket{u_2 u_3} q_{2\mathcal{I}} q_3^{\mathcal{I}}}{\sigma_{23}}\right)^{\!2},\qquad \text{'gluon'},\\      
       \;\frac{q_{2\mathcal{I}} q_3^{\mathcal{I}}}{\sigma_{23}} \frac{\text{det}H^{[23]}_{[32]}}{\braket{u_3 u_2}} = \text{det}^{\prime} H \frac{\braket{u_2 u_3} q_{2\mathcal{I}} q_3^{\mathcal{I}}}{\sigma_{23}}\;\;, \qquad\qquad\quad\text{'gravitino'}.
       \end{cases}
\end{equation*}
It turns out that the amplitude is the same as the helicity-conserving case \eqref{ComptAmpl}, which is good for consistency. Because a PT factor for massless particles needs to have at least one particle of each helicity, the 'gluon' case cannot flip the helicity, but it is possible for a 'gravitino':
\begin{equation*}
\text{det} G^{[34]}_{[12]} = G_{13} G_{24} - G_{14} G_{23} = \frac{\epsilon_1^A \epsilon_{4 A} q_{2\mathcal{I}} q_3^{\mathcal{I}}}{\sigma_{14} \sigma_{32}} = \frac{q_{2\mathcal{I}} q_3^{\mathcal{I}}}{\epsilon_2^A \epsilon_{3 A}} \text{det} H^{[34]}_{[12]}.
\end{equation*}
Therefore, going from spin 2 to 3/2 just lowers the power of $[23]$ in \eqref{sameHelAmpl}, but it also lowers the power in the gravitational coupling, in contrast to the helicity-preserving case. The same is true for the massive particle:
\begin{equation*}
 \label{ComptAmpl1}
 M_4^{++} = \frac{\langle \mathbf{1} \mathbf{4} \rangle^{2s} \, [2 3]^{2h} }{(2p_1 \!\cdot\! k_2 ) (2k_2 \!\cdot\! k_3) (2k_2 \!\cdot\! p_4)},\quad 3/2 \leq h,s \leq 2\;.
 \end{equation*}
  Finally, one can consider both, the massless and the massive, particles as 'gluons' leading to replace det$^{\prime}G$ with a PT factor for two incoming and two outgoing particles, i.e. the situation is now described by the Coulomb branch of SYM for the scattering between 'gluons' and 'W bosons' (see section \ref{consChk}): 
 \begin{equation*}
 <W g^{+} g^{+} \bar{W}\!> = \frac{\braket{ \mathbf{1} \mathbf{4} }^2 [2 3]}{(2 k_2 \!\cdot\! p_4) \braket{2 3}}, \quad <W g^{+} g^{-} \bar{W}\!> = \frac{(\langle \mathbf{4} 3 \rangle \, [\mathbf{1} 2] - \langle \mathbf{1} 3 \rangle \, [\mathbf{4} 2])^2}{(2 k_2 \!\cdot\! p_4) (2 k_2 \!\cdot\! k_3)},
 \end{equation*}
 which is consistent with corresponding results in \cite{Craig:2011ws, Cachazo:2018hqa, Albonico:2023}.\\

For gravitational Compton scattering on a Kerr black hole, to make contact with the results in \cite{Guevara:2018wpp, Aoude:2020onz, Bautista:2021wfy, Bautista:2022wjf}, the kinematic variables are now taken to be the same as in \cite{Bautista:2021wfy, Bautista:2022wjf}.
The black hole is treated as a massive spin particle with mass $M$ and a spin vector $a^{\mu} = s^{\mu} / M = (0, a_x, a_y, a_z)$. The incoming/outgoing momenta of the scattered particle are still labeled as $k_2/k_3$ and of the black hole as $p_1/p_4$. In the chosen reference frame the black hole is taken to be initially at rest and the scattering process to happen in the $x - z$ plane. This leads to the following explicit representation of the kinematic variables:  
  \begin{equation*}
\label{kinematics}
    \begin{split}
    p_1^{\mu} & =(M,0,0,0),\\
k_2^{\mu} & =\hbar \omega(1,0,0,1),\\
k_3^{\mu} & =\gamma \, \hbar \omega(1,\sin\theta,0,\cos\theta),
\gamma = 1 / {(1+2\frac{\hbar \omega}{M}\sin^2\!\frac{\theta}{2})},\\
p_4^{\mu} & =p_1^{\mu}+k_2^{\mu}-k_3^{\mu}\,.
    \end{split}
\end{equation*}
Here $\theta$ is the scattering angle and the value of $\gamma$ is determined by the on-shell condition $p_4^2 = M^2$ for the outgoing massive momentum. It should be kept in mind that this setup violates the standard notation mentioned at the beginning of the section, by reversing the incoming momenta $p_1$ and $k_2$.\\

It will be convenient to introduce a basis of spinor helicity variables in terms of $\theta$ to represent the momentum and the polarization matrices of the outgoing massless particle:
\begin{align*}
      &[3| = \sqrt{2 \gamma \hbar \omega} \left(\cos\frac{\theta}{2}, \sin\frac{\theta}{2} \right), && [\mu| = \left(- \! \sin\frac{\theta}{2}, \cos\frac{\theta}{2} \right)\,,\nonumber\\
      &|3\rangle = \sqrt{2 \gamma \hbar \omega} \left(\cos\frac{\theta}{2}, \sin\frac{\theta}{2} \right), &&|\mu\rangle = \left(- \! \sin\frac{\theta}{2}, \cos\frac{\theta}{2} \right)\,,\nonumber\\
      &k_{3\,\alpha \dot{\alpha}} = |3\rangle_{\alpha} [3|_{\dot \alpha}\,, &&\nonumber\\
      &\epsilon^{+}_{3\,\alpha \dot{\alpha}} = \sqrt{2} \frac{|\mu \rangle_{\alpha} [3|_{\dot \alpha}}{\langle \mu 3\rangle}\,, &&\epsilon^{-}_{3\,\alpha \dot{\alpha}} =  \sqrt{2}\frac{|3 \rangle_{\alpha} [\mu|_{\dot \alpha} }{[3 \mu]} \,,
\end{align*}
where $|\mu\rangle$ and $[\mu|$ serve as reference spinors. The incoming massless states associated to $k_2$ are recovered by setting $\theta=0$. Other choices for the reference spinors are allowed. Actually, the reference spinor $|\mu\rangle$ in $\epsilon^{+}_3$ and $\epsilon^{+}_2$ could be replaced by $|2\rangle$ and $|3\rangle$, respectively, and $[\mu|$ in $\epsilon^{-}_3$ and $\epsilon^{-}_2$ by $[2|$ and $[3|$, respectively, a choice which will be made presently.\\
  
  In this kinematics the classical limit ($s \rightarrow \infty, \gamma \rightarrow 1$) of $M_4^{+-}$ in \eqref{ComptAmpl} can be written as \cite{Guevara:2018wpp, Aoude:2020onz, Bautista:2022wjf}
  \begin{equation}
  \label{GravComptAmpl}
  \begin{split}
  M_4^{+-} &= \frac{M^2 (\cos\frac{\theta}{2})^{2h}}{\sin^2 \frac{\theta}{2}} \exp[(k_3 - k_2 + 2w)\cdot a] \\
  &= \frac{M^2 (\cos\frac{\theta}{2})^{2h}}{\sin^2 \frac{\theta}{2}} \exp\left[-\frac{\hbar \omega}{M} \left(2 a_z\sin^2 \frac{\theta}{2} - a_x \sin \theta + 2(a_x - ia_y) \tan\frac{\theta}{2}\right) \right]\;,
  \end{split}
  \end{equation}
  where the vector $w$ is defined as $w = - p_1\!\cdot\! k_3 \frac{|3\rangle[2|}{\langle 3 | p_1 | 2]} = - \frac{p_1 \!\cdot k_3}{p_1 \!\cdot \epsilon_3^{-}} \epsilon_3^{-}$.\\
  
  Amplitudes \eqref{GravComptAmpl} and \eqref{ComptAmpl} for $s > h$ exhibit a spurious pole for $\langle 3|p_1|2 \rangle = 2 M \hbar \omega \cos \frac{\theta}{2}$ that stems from the vector $w$ and needs to be remedied with help of contact terms \cite{Bautista:2022wjf}. However, the issue of contact terms will not be considered here. But it is intriguing to note that the spurious pole is missing in the reverse-helicity scattering amplitude $M_4^{++}$ and that indeed according to \eqref{auxHel} $w$ can be rewritten as $w = \frac{\langle k_2 1 4\rangle}{2\langle 1 2 3 4 \rangle}|3\rangle[2| = -\!\braket{u_2 u_3}\!/(2\sigma_{23})|3\rangle[2|$ which vanishes for massless particles when flipping the helicity, that is, from the point of view of the ambitwistor string model, the appearance or absence of the spurious pole is directly related to solutions of the polarized scattering equations.\\
  
  In summary, although this section has not provided any new results (besides maybe the gravitino amplitudes), it was interesting to show generally and in the context of a black hole how, starting from a Lagrangian, the ambitwistor string models with a single amplitude \eqref{massComptonAmpl} can describe multiple aspects of the Compton scattering of a massless particle of various spin hitting a massive target particle with higher spin.\\

\section{Summary and Outlook}
  \label{Summary} 
  This work presented an anomaly-free extension to the massive ambitwistor string found in \cite{Geyer:2020, Albonico:2022, Albonico:2023} as a unified model for supergravity and the Coulomb branch of SYM. The all-multiplicity tree amplitudes were shown to have the expected massless limit, to include EYM amplitudes for multiple gluon traces, and to have proper unitary factorization. The all-multiplicity one-loop amplitude was calculated and established to have unitary factorization for separating and non-separating degeneration. At the one-loop level the cosmological constant is zero. It was argued and referred to \cite{Kunz:2025} for verification that this extends to all loop levels. Then Compton scattering of a massless particle against a massive target was used to obtain scattering amplitudes for both cases of conserved and reversed helicity. They were compatible with results in the literature. One surprise score might be counted for the non-vanishing reverse-helicity amplitude of gravitino scattering.\\
   
   Future work would include the calculation of higher loop amplitudes together with the demonstration of proper factorization \cite{Kunz:2025}. Because the model will be IR-divergent when massless particles like the graviton and photon are included, IR-renormalization will be required. In order to do that correctly, a string field theory might need to be developed. It would be interesting to see how model features would be affected by renormalization and whether for instance the cosmological constant could pick up a non-vanishing value. Further, the $\mathcal{N} = 8$ supergravity spectrum with SU(4) $\!\times\!$ SU(4) R-symmetry should be studied for various symmetry breakdown scenarios to get more realistic spectra\footnote{Note that the $\mathcal{N} = 8$ supergravity spectrum contains exactly the spin $\frac{1}{2}$ content for all the fermions of the standard model and also for 8 massive gravitinos, a fact that is well known \cite{Meissner:2014joa, Meissner:2018gtx}.}.\\
   
   And, of course, the origin of all the auxiliary spinors in the action \eqref{massSugraAction} is still a mystery. They are excluded from the physical spectrum and the question arises whether they materialize as a ghost system used for mathematical convenience or whether they emanate from compactifying extra spinor space dimensions (in which case there still might not be a mandatory necessity to relate them to extra spacetime dimensions). More knowledge in this regard might also reveal the exact nature of the unknown six bosonic degrees of freedom that make the model anomaly-free.\\

\bibliography{TwistorString}
\end{document}